\documentclass[b5paper,twoside]{jpconf}  
\usepackage{graphicx}

\usepackage{bm}

\newcommand{\eqb}{\begin{eqnarray}}
\newcommand{\eqe}{\end{eqnarray}}

\begin{document}

\title[Pulsar Wind]{Of Winds and Waves}

\author[Kirk \& Arka]{John G. Kirk$^1$ and Ioanna Arka$^2$}

\address{$^1$Max-Planck-Institut f\"ur Kernphysik, Postfach 10 39 80, 69029 Heidelberg, Germany}
\address{$^2$Institute de Planetologie et d'Astrophysique, University of Grenoble, Grenoble, France}

\ead{john.kirk@mpi-hd.mpg.de} 

\begin{abstract}
Recent work on the properties of superluminal waves in pulsar winds is 
summarized. It is speculated that these waves play an important role in 
 the termination shock that divides the wind from the surrounding nebula.
\end{abstract}
\section{Introduction}
In addition to pulses of radiation, rotation-powered pulsars are
thought to emit a wind that powers the diffuse radiation or {\em
  pulsar wind nebul\ae\/} (PWN) observed around many of them in the
radio, optical, X-ray and gamma-ray bands.  Synchrotron emission and
inverse Compton scattering are the mechanisms most likely to be at
work in PWN, and it has been evident for over half a century that not
only the energy, but also both the relativistic electrons (or
positrons) and the magnetic fields in these nebul\ae\ 
must be supplied by the central star\cite{piddington57}.

The basic idea is that the rotation of the neutron star couples to the
wind through the strong magnetic field anchored in its crust
\cite{pacini67,reesgunn74}. As well as electromagnetic fields
oscillating at the rotation frequency of the pulsar, the wind contains
a DC (phase-averaged) component of the magnetic field, electrons and
positrons created by cascades in the magnetosphere and, perhaps, a
relatively small number of electrons and ions extracted from the
stellar surface. As it propagates away from the star, the ram pressure
of the wind decreases, and, roughly where it equals the ambient
pressure, a {\em termination shock} is formed.  Rees \& Gunn
\cite{reesgunn74} simply assumed that the waves are absorbed at the
termination shock, leaving behind the energized particles and magnetic
fields that fill the nebula. Kundt \& Krotscheck
\cite{kundtkrotscheck80}, however, noted that 
electromagnetic waves could also be reflected by
the shock and build up inside it. 
This raises an 
important question concerning the structure of the termination shock
and the wind it encloses. Because the waves have large amplitude, they
interact strongly with each other and with the particle component. Rather 
than just a linear superposition of DC fields, particles, 
outward propagating waves and 
reflected waves, one must, therefore, look for self-consistent
solutions, containing all of these, 
that match the outer boundary conditions.

\section{The $\bm{\sigma}$ problem}

Estimates of the energy density in magnetic field at 
the light cylinder $r_{\rm L}=c/\omega$ (with $\omega$ the angular 
frequency of the pulsar) suggest that 
the wind is energetically dominated at launch by the waves
and/or DC fields. On the other hand, outside the termination shock the
particle pressure is comparable or larger than the pressure exerted by
the fields (for reviews, see \cite{gaenslerslane06,kirketal09}). The
implied conversion of electromagnetic into kinetic energy is quite
natural in many situations, for example, when an MHD outflow is
collimated and accelerated 
into a jet.  However, this kind of conversion does not
appear possible in a pulsar wind, where the poloidal flux 
threads the neutron star \cite{kirketal09}. 
In such a case, the MHD equations
suggest the flow remains radial and its (supermagnetosonic) speed stays
constant until the termination shock is reached. 

This, then, is the \lq\lq $\sigma$ problem\rq\rq\ 
(the ratio of Poynting to kinetic energy
flux is conventionally denoted by $\sigma$): if the pulsar wind
can be described by the equations of ideal MHD, there is no way in
which a large value of $\sigma$ near the 
star can be reduced to unity or below before the termination shock is
reached. Furthermore, if the termination shock is simply a
discontinuity that obeys the usual jump conditions 
in an otherwise ideal MHD flow, the downstream plasma
remains magnetically dominated with large $\sigma$.

\section{Dissipation in current sheets}
To solve this problem, we obviously need to go beyond the ideal MHD
description \cite{lyutikov10}.  One possibility, suggested by Coroniti \cite{coroniti90}
and Michel \cite{michel94}, 
is to allow for dissipation in a current sheet
embedded in a wind that is otherwise a cold, MHD flow. This is the
\lq\lq striped wind\rq\rq\ picture.   
Conversion of the energy flux from fields to particles by 
dissipation in the sheets causes the 
flow to 
accelerate \cite{lyubarskykirk01,drenkhahn02,lyubarsky10} 
--- in the simplest model the bulk Lorentz factor 
rises as $\Gamma\propto r^{1/2}$. Consequently, time-dilation reduces
the effective dissipation rate seen in the lab.\ frame. 
It is possible to place upper and
lower limits on the dissipation rate 
 in order to 
constrain the
characteristic radius at which the oscillating components of the
fields are annihilated.
The results depend on the pair-loading of the flow,
described by the parameter 
\eqb 
\mu&=&\frac{L}{\dot{M}c^2} 
\eqe 
where
$L$ is the luminosity and $\dot{M}$ the mass-loss rate. For the Crab,
dissipation occurs before the termination shock is reached only if 
$\mu <10^4$, a value somewhat lower than
conventional estimates \cite{kirkskjaeraasen03}.  

Embedded current sheets also change our picture of the termination
shock.  On the basis of an analytical model and 1D PIC simulations,
Lyubarsky \& Liverts \cite{lyubarskyliverts08} and P\'etri \&
Lyubarsky \cite{petrilyubarsky07} suggested that an MHD shock sweeping
through the cold portions of the striped wind would drive reconnection
in the embedded current sheets and dissipate a substantial fraction of the 
magnetic energy,  provided the resulting heated electrons
were unmagnetized. For $\sigma\gg1$, this is equivalent to the condition
\eqb
\mu&>&a
\label{rcritcondition1}
\eqe where the strength parameter $a$ of the incoming wave (striped
wind) is the ratio of the quiver frequency of the electrons $eB/mc$ to
the angular frequency $\omega$ of the wave. (see \cite{kirk10}).%
\footnote{There are several ways to define this important parameter.
In the striped wind, where $\left|B\right|$ is phase-independent
and $\sigma\gg1$, there
is no ambiguity. 
For the superluminal waves discussed in section~\ref{superluminalwaves} we
  take $a=eE_{\rm max}/mc\omega$, where $E_{\rm max}$ is the amplitude
  of the oscillating electric field. A covariant, gauge-independent
  definition for vacuum waves is given by Heinzl \& Ilderton
  \cite{heinzlilderton09}} 
Since $a$ decreases as $1/r$ in spherical
geometry, (\ref{rcritcondition1}) amounts to a lower limit on the
radius at which such a process could operate: \eqb r&>&r_{\rm crit}
\label{rcritcondition2}\\
&\approx&3.4\times10^{10} r_{\rm L} L_{38}^{1/2}/\mu
\nonumber
\eqe
where $\left(L_{38}/4\pi\right)\times10^{38}\textrm{erg/s}$ is the 
luminosity per unit solid angle carried by the wind.
 
On the other hand, Sironi \& Spitkovsky \cite{sironispitkovsky11}
recently considered driven reconnection in the striped wind using 2D
and 3D PIC simulations. Although their investigations were confined to
the range $0.05<r/r_{\rm crit}<4$, they found substantial dissipation
in all cases. They also noted the appearance of high-energy particles
accelerated by the first-order Fermi mechanism, and, on the basis of a
comparison of the nonthermal particle spectrum with the observed
synchrotron spectrum, suggested that, in the case of the Crab, the
termination shock should be located where $r\approx r_{\rm
  crit}/3$. If this interpretation is correct, the radius of the
termination shock inferred from X-ray observations ($\sim0.1\,$pc)
implies $\mu\sim50$. This is much smaller than conventionally
estimated, implying a very high pair loading and a mildly relativistic
$\Gamma\sim10$ outflow. An approximately spherically symmetric wind
with these parameters would vastly overpopulate the nebula with
electrons and positrons. However, if we interpret the wind parameters
and, therefore, the radius of the termination shock, as latitude
dependent, it cannot be ruled out that some part of the wind, for
example the equatorial plane, contains a relatively dense, slow
outflow of the type suggested.

\section{Charge starvation}

As well as giving rise to dissipation in current sheets, non
(ideal-)MHD effects can also manifest themselves via charge
starvation, which can arise, for example, when the required currents
demand relativistic motion of the current carriers
\cite{kirkmochol11,kirkmochol11a}. A relativistic two-fluid
(electron-positron) approach can be used to model this effect. In the
simplest case, with cold fluids and charge neutrality, both subluminal
and superluminal nonlinear waves can be found \cite{kirk10}. 

The subluminal wave resembles the striped wind, except that the current
sheet is replaced by a static shear, with $\left|\vec{B}\right|^2$ constant
(i.e., a circularly polarized wave). The phase profile of the shear can be 
chosen arbitrarily, but it is monochromatic in the simplest 
case of constant density. 
It is interesting to note that,
as for the striped wind, the non-MHD effects cause also
this solution to accelerate with radius, in this case with
$\Gamma\propto r$, despite the fact that no dissipation process is 
involved \cite{kirkmochol11}. 
However, at least for the monochromatic wave, acceleration starts 
relatively far from the pulsar, where
\eqb
\sigma&>&a
\label{rstarvecondition1}
\eqe
corresponding to 
\eqb
r&>&\Gamma r_{\rm crit}
\label{rstarvecondition2}
\eqe

The solutions with superluminal phase speed resemble vacuum electromagnetic
waves, in the sense that the displacement current plays a crucial role.
The fields are transverse, and satisfy 
$\left|\vec{E}\right|>\left|\vec{B}\right|$. The waves and particles
are not locked together in the way they are in the subluminal solutions, which
makes them more promising candidates for matching to a boundary condition 
at the termination shock. 
There have been extensive investigations of these
waves in the literature (see \cite{arkakirk11} and references therein). They
propagate only when the density falls below a critical value. In spherical 
geometry, the corresponding condition on the radius coincides with
(\ref{rcritcondition2}). Because of this, they cannot be launched close to 
the pulsar, and can only be realized if a striped-wind-like (subluminal) 
solution converts into a superluminal one either spontaneously, or because
it is forced to do so by the boundary condition imposed by the termination 
shock. In this case, the conversion process and subsequent radial evolution
and damping of the mode should more properly be regarded as an integral  
part of the \lq\lq termination shock\rq\rq\ structure.
 
\begin{figure}
\includegraphics[width=\textwidth]{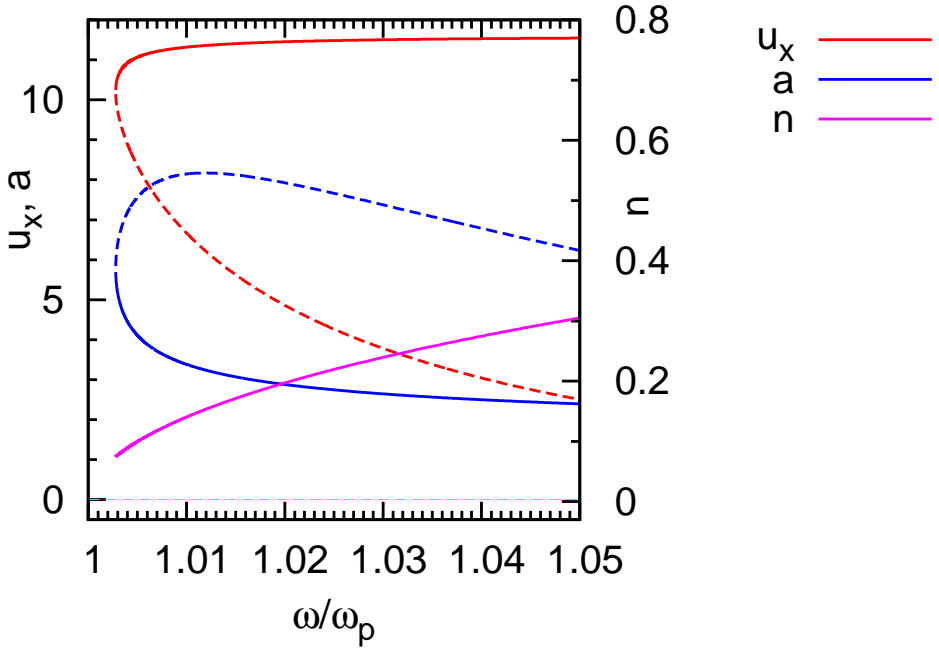}
\caption{\label{test_circ2}%
The radial four-velocity of the fluid, the strength parameter $a$, and 
the refractive index $n$ of 
circularly polarized superluminal waves close to the cut-off
frequency for $\mu=12$, $\sigma=3$.}
\end{figure} 
\section{Superluminal waves}
\label{superluminalwaves}
 A first step towards understanding the role of superluminal waves is 
to determine the propagation characteristics of those waves to which 
a given striped (or other subluminal) wind can convert. This amounts to solving 
the two-fluid analogue of the shock jump conditions. Here we present 
a brief summary of some recent work on this topic --- a full description 
can be found in \cite{arkakirk11}. 

Circularly polarized modes with zero phase-averaged field are the simplest to 
analyze, since the jump conditions can be solved analytically
\cite{kirk10}. Figure~\ref{test_circ2} illustrates the properties of these 
waves for parameters $\mu=12$, $\sigma=3$ (and, therefore, $\Gamma=3$).
Although not in the range expected in pulsar winds, this parameter set 
 reveals the mode structure near the cut-off frequency particularly clearly. 
The figure shows that for each value of the frequency $\omega$ in 
the lab.\ frame, measured in units of the proper plasma frequency 
$\omega_{\rm p}$, two solutions exist that carry the prescribed particle, 
energy and momentum fluxes: a strong wave through which the plasma streams 
relatively slowly (dashed line) and a weaker one  (solid line) 
through which the plasma streams rapidly ($u_{x}$ is the four velocity
along the propagation direction). Transition to the the weak wave
involves almost complete annihilation of the incoming field 
energy, leading to $u_x\approx \mu$. 
The refractive 
indices of these modes coincide and correspond to superluminal 
phase speed $n<1$. Because of the finite wave amplitude, the lab.\ frame 
frequency $\omega$ exceeds the proper plasma frequency $\omega_{\rm p}$, which
equals the cut-off frequency for linear waves.  

\begin{figure}
\includegraphics[width=\textwidth]{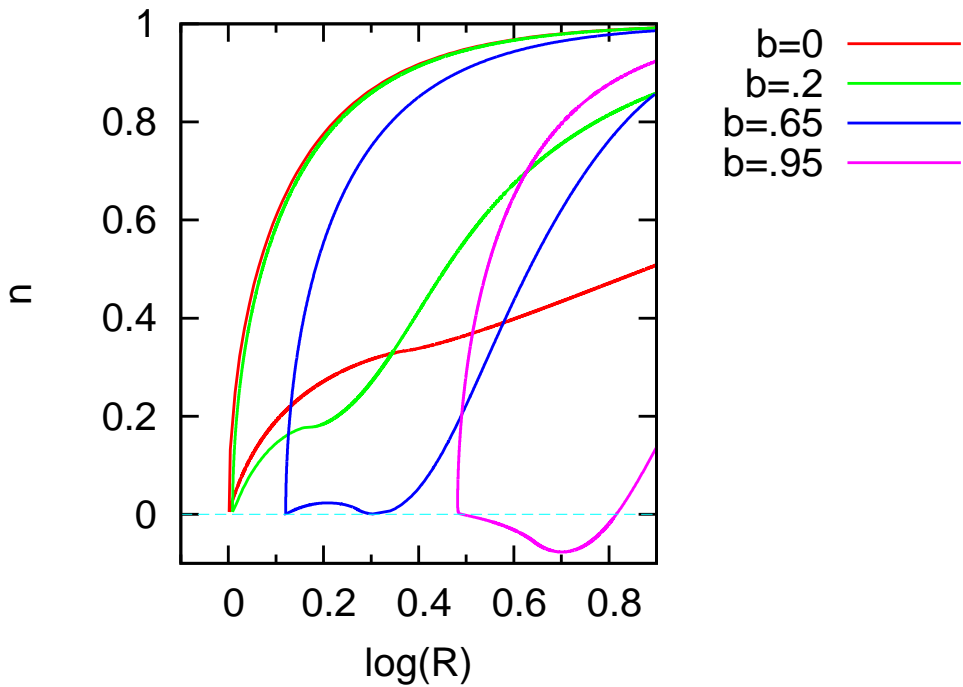}
\caption{\label{linear2}%
The refractive indices 
of linearly polarized superluminal waves close to the cut-off
frequency for $\mu=10100$, $\sigma=100$ as a function of radius normalized to 
$r_{\rm crit}$ (see Eq.~\ref{rcritcondition2}), for different values of the 
phase-averaged magnetic field}
\end{figure} 
\begin{figure}
\includegraphics[width=\textwidth]{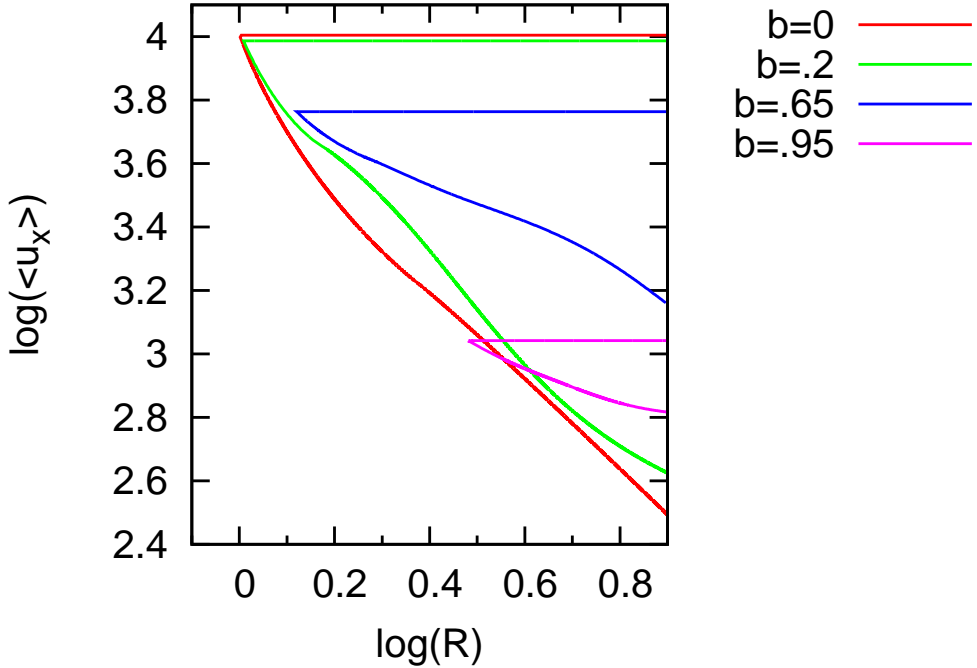}
\caption{\label{linear3}%
The phase-averaged radial four-speed of the electron and positron fluids
in the linearly polarized superluminal waves shown in Fig.~\ref{linear2}.}
\end{figure} 
Linearly polarized waves are much more complicated, requiring a
numerical solution to the jump conditions. Nevertheless, the
properties illustrated in Fig.~\ref{test_circ2} apply also to these
modes.  Two solutions are available for each set of \lq\lq
upstream\rq\rq\ parameters that characterize the corresponding
subluminal solution. This is shown in Fig.~\ref{linear2}, which
differs from Fig.~\ref{test_circ2} in two ways. Firstly, we show
the refractive index not as a function of frequency, but of radius in
the pulsar wind, normalized to $r_{\rm crit}$ given in
Eq.~(\ref{rcritcondition2}). In this representation, the refractive
indices of the two solutions no longer coincide, since the proper
plasma frequencies associated with each solution, 
differ although the lab.\ frame densities do not.
Secondly, we show solutions for four different values of the
phase-averaged magnetic field, given by the parameter
$b=\left<\vec{B}\right>/\left<B^2\right>^{1/2}$.  It can be seen that
stronger DC fields push the cut-off radius further out from the
pulsar. Furthermore, stronger fields display solutions with negative
refractive index, corresponding to phase velocities directed radially
inwards. Figure~\ref{linear3} shows the phase-averaged
four-velocity of the plasma stream in the radial direction. Note that
the weak wave has $\left<u_x\right>\approx\mu$ only for zero averaged
field.  For finite values of the DC field, only a fraction of the
incoming Poynting flux is carried by the oscillating fields and 
can be annihilated.

\section{Summary}

Because they are powerful enough to evacuate a relatively large cavity,
pulsar winds provide an environment in which electromagnetic fields dominate 
the dynamics, and, especially in the outer regions, 
non-MHD effects are crucial. Within the critical radius given by 
(\ref{rcritcondition2}), energized electrons and positrons,
which have Lorentz factors $\sim\mu$, can be regarded
as magnetized, since their relativistic Larmor radius, $\mu mc^2/eB$ 
(which in this case equals the effective inertial length $c/\omega_{\rm p}$) 
is smaller than the length scale $c/\omega$. Here
the superluminal modes cannot propagate, because 
the plasma is sufficiently dense to screen out 
electromagnetic waves reflected from the termination shock. 
In this region, 2D and 3D PIC simulations 
indicate that dissipation could be driven by a standing shock front
\cite{sironispitkovsky11}. However, in the case of isolated pulsars,
$r_{\rm crit}$ lies well inside the point of pressure balance with the 
external medium, making it doubtful that such a shock could be supported.

Outside $r_{\rm crit}$ energized particles are unmagnetized.
Driven reconnection at the termination shock may still
occur \cite{petrilyubarsky07}, but superluminal waves may also play an
important role. In section~\ref{superluminalwaves} we have briefly
summarized some recent work \cite{arkakirk11} on the properties of
these waves for parameters thought appropriate in pulsar winds, and
sketched their possible role.  Although the physics of these two
scenarios remains to be investigated in detail, it seems likely that
there will be observable implications.  Indeed, current sheets have
been proposed as the source of both the pulsed high-energy emission
\cite{petri11,petridubus11} as well as the the recently observed
flares in very high energy gamma-rays from the Crab
\cite{uzdenskyetal11,ceruttietal11}.

\bibliographystyle{iopart-num}
\section*{References}

\begin{thebibliography}{10}
\expandafter\ifx\csname url\endcsname\relax
  \def\url#1{{\tt #1}}\fi
\expandafter\ifx\csname urlprefix\endcsname\relax\def\urlprefix{URL }\fi
\providecommand{\eprint}[2][]{\url{#2}}

\bibitem{piddington57}
{Piddington} J~H 1957 {\em Australian Journal of Physics\/} {\bf 10} 530--+

\bibitem{pacini67}
{Pacini} F 1967 {\em Nature\/} {\bf 216} 567--+

\bibitem{reesgunn74}
{Rees} M~J and {Gunn} J~E 1974 {\em \mnras\/} {\bf 167} 1--12

\bibitem{kundtkrotscheck80}
{Kundt} W and {Krotscheck} E 1980 {\em \aap\/} {\bf 83} 1--21

\bibitem{gaenslerslane06}
{Gaensler} B~M and {Slane} P~O 2006 {\em Ann. Rev. Astron. Astrophys\/} {\bf
  44} 17--47 (\textit{Preprint} \eprint{arXiv:astro-ph/0601081})

\bibitem{kirketal09}
{Kirk} J~G, {Lyubarsky} Y and {Petri} J 2009 {The Theory of Pulsar Winds and
  Nebulae} {\em Astrophysics and Space Science Library\/} ({\em Astrophysics
  and Space Science Library\/} vol 357) ed {W~Becker} pp 421--+

\bibitem{lyutikov10}
{Lyutikov} M 2010 {\em \mnras\/}  581--+ (\textit{Preprint} \eprint{0911.0324})

\bibitem{coroniti90}
{Coroniti} F~V 1990 {\em \apj\/} {\bf 349} 538--545

\bibitem{michel94}
{Michel} F~C 1994 {\em \apj\/} {\bf 431} 397--401

\bibitem{lyubarskykirk01}
{Lyubarsky} Y and {Kirk} J~G 2001 {\em \apj\/} {\bf 547} 437--448
  (\textit{Preprint} \eprint{arXiv:astro-ph/0009270})

\bibitem{drenkhahn02}
{Drenkhahn} G 2002 {\em \aap\/} {\bf 387} 714--724 (\textit{Preprint}
  \eprint{arXiv:astro-ph/0112509})

\bibitem{lyubarsky10}
{Lyubarsky} Y 2010 {\em \apjl\/} {\bf 725} L234--L238 (\textit{Preprint}
  \eprint{1012.1411})

\bibitem{kirkskjaeraasen03}
{Kirk} J~G and {Skj{\ae}raasen} O 2003 {\em \apj\/} {\bf 591} 366--379
  (\textit{Preprint} \eprint{arXiv:astro-ph/0303194})

\bibitem{lyubarskyliverts08}
{Lyubarsky} Y and {Liverts} M 2008 {\em \apj\/} {\bf 682} 1436--1442
  (\textit{Preprint} \eprint{0805.0085})

\bibitem{petrilyubarsky07}
{P{\'e}tri} J and {Lyubarsky} Y 2007 {\em \aap\/} {\bf 473} 683--700

\bibitem{kirk10}
{Kirk} J~G 2010 {\em Plasma Physics and Controlled Fusion\/} {\bf 52} 124029--+
  (\textit{Preprint} \eprint{1008.0536})

\bibitem{heinzlilderton09}
{Heinzl} T and {Ilderton} A 2009 {\em Optics Communications\/} {\bf 282}
  1879--1883 (\textit{Preprint} \eprint{0807.1841})

\bibitem{sironispitkovsky11}
{Sironi} L and {Spitkovsky} A 2011 {\em ArXiv e-prints\/} (\textit{Preprint}
  \eprint{1107.0977})

\bibitem{kirkmochol11}
{Kirk} J~G and {Mochol} I 2011 {\em \apj\/} {\bf 729} 104--+ (\textit{Preprint}
  \eprint{1012.0307})

\bibitem{kirkmochol11a}
{Kirk} J~G and {Mochol} I 2011 {\em \apj\/} {\bf 736} 165--+

\bibitem{arkakirk11}
{Arka} I and {Kirk} J~G 2011 {\em ArXiv e-prints\/} (\textit{Preprint}
  \eprint{1109.2756})

\bibitem{petri11}
{P{\'e}tri} J 2011 {\em \mnras\/} {\bf 412} 1870--1880 (\textit{Preprint}
  \eprint{1011.3431})

\bibitem{petridubus11}
{P{\'e}tri} J and {Dubus} G 2011 {\em \mnras\/}  1193--+ (\textit{Preprint}
  \eprint{1104.4219})

\bibitem{uzdenskyetal11}
{Uzdensky} D~A, {Cerutti} B and {Begelman} M~C 2011 {\em \apjl\/} {\bf 737}
  L40+ (\textit{Preprint} \eprint{1105.0942})

\bibitem{ceruttietal11}
{Cerutti} B, {Uzdensky} D~A and {Begelman} M~C 2011 {\em ArXiv e-prints\/}
  (\textit{Preprint} \eprint{1110.0557})

\end{thebibliography}
\providecommand{\newblock}{}

\end{document}